\newcommand\bm[1]{\mbox{\boldmath$#1$}}

\documentclass[12pt,a4paper]{article}
\usepackage{color}
\usepackage{amsfonts,amssymb}

\begin{document}

\title{An approximate global solution of Einstein's equations for a rotating two layers star}
\author{A.\ Molina${}^{1,2}$ and E.\ Ruiz${}^3$\\[.5ex]
${}^1$\emph{Dep. de F\'\i sica Qu\`antica i Astrof\'\i sica 
}
\\
${}^2$\emph{Institut de Ci\`encies del Cosmos (ICCUB)}
\\
\emph{Universitat de Barcelona},
\\
\emph{Mart\'{\i} Franqu\`es 1, 08028 Barcelona, Spain}
\\
${}^3$\emph{Instituto  Universitario de F\'\i sica Fundamental y Matem\'aticas},
\\
\emph{Universidad de Salamanca},
\\
\emph{Plaza de la Merced s/n, 37008 Salamanca, Spain}.
}
\maketitle

\date{}

\begin{abstract}
We obtain an approximate global stationary and axisymmetric solution of Einstein's equations which can be thought as a simple two layers star model: a self-gravitating  ball built up by two layers of perfect fluid having different linear equation of state moving  in a rigid motion pattern. Using the post-Min\-kows\-kian formalism (weak-field approximation) and considering rotation as a perturbation  (slow-rotation approximation), we find approximate interior and exterior (asymptotically flat) solutions to this problem in harmonic coordinates. Interior and exterior solutions are matched, in the sense described by Lichnerowicz, on the
surfaces of constant pressure, to obtain a global solution. 

\noindent
PACS number(s) 04.40.Nr, 04.20.Jb
\end{abstract}

\section{Introduction}
Stellar models for rotating stars are built matching an interior space-time describing the source and 
the exterior space-time that encloses it. A good candidate for the interior solution could correspond
to a stationary axisymmetric perfect fluid without extra symmetries and which admit a zero
pressure surface. 

Summing to the difficulty of finding suitable interiors, there are the ones arising from the matching with the asymptotically flat exterior. It is an overdetermined problem \cite{Mars-Seno} so in general we can not find an exterior that matches a given interior. Such seems to be the case for Wahlquist,
\cite{Wahlquist} where the derivations of the impossibility of matching it with an asymptotically flat exterior  come from the analysis of the shape of its surface and involve approximations.\cite{BFMP},\cite{CuMMR}

We have built  approximate solutions of the Einstein equations which describes the gravitational field
inside a ball of perfect rotating fluid. We have studied  fluids with a linear EoS \cite{CR},\cite{CMMR},\cite{CuGMR} or with a polytropic equations of state,\cite{MMR}  and also differentially rotating fluids with a linear EoS\cite{MR}. We have solved all these problem in harmonic coordinates. 

Within the class of analytical solutions for these problems there is a very known approach due to Hartle, 
\cite{Hartle}, \cite{Reina-Vera}, \cite{Marc-Vera} where also an approximation scheme to find solutions is used.
This approach begin  with a spherical symmetric exact solution in the null rotation limit and a slow rotation perturbation method is implemented. The main problem for this approach is that there is no an exact analytical solution and at the end a numerical approach is required, another problem is the difficulty of the Hartle approach  to obtain higher order solutions in the slow rotation parameter $\omega$. Nevertheless higher orders of $\omega$ have been obtained.\cite{Yagi} 

Our approximation scheme  consist in a slow rotation approximation on a post-Minkowskian algorithm. We introduce two dimensionless parameters. One, $\lambda$, measures the strength of the gravitational field, the other,
$\Omega$, measures the deformation of the matching surface due to fluid rotation. With this two parameters method higher order analytic solutions can be easily obtained except for the polytropic EoS where the results can be only implicitly obtained in terms of the Lane-Emdem equation and its integrals.  

If there is no rotation ($\Omega=0$), we are faced
to the post-Minkowskian perturbation of the gravitational field for a spherically symmetric mass distribution. On the other hand, Newtonian deformation of the source due to rotation is included in the first order $\lambda$ terms up to some order in the rotation parameter.

In our previous articles we have computed  terms of order less than or equal to $\Omega^5$ and  $\lambda^{5/2}$; that is, we have gone beyond a simple linear analysis but no so far to compute strong non-linear effects. However, since the algorithm is implemented by an algebraic computational program, our results  can easily be enhanced, if so desired, by going farther in the approximation scheme.

In this paper we are going to build a solution up to first order in the parameter $\lambda$  for a rotating ball of perfect fluid with a linear EoS $\mu+(1-n_1)p=\mu_1$  (core) surrounded by a co-rotating thick layer with EoS density $\mu+(1-n_2)p=\mu_2$ and we will match this interior solution with an asymptotically flat vacuum space-time. 
We have computed the global metric with terms of order $\lambda^{5/2}$ and $\Omega^3$, nevertheless in this paper, due to the length of the expressions obtained we  only give the terms up to the order $\lambda^{3/2}$ and $\Omega^3$. 

In this paper we have included some repetitive content, for instance in the next section, with respect to our previous work but also there are some incompleteness that can be solved by looking at our previous articles. \cite{CR},\cite{CMMR},\cite{CuGMR}

\section{The metric and energy--momentum tensor}

The solution we are looking for is a stationary, 
axisymmetric and asymptotically flat space-time that admits a global system of spherical-like coordinates $\{t,r,\theta,\varphi\}$.

Our coordinates are adapted to the space-time symmetry, $\bm\xi=\partial_t$ and
$\bm\zeta=\partial_\varphi$, which are respectively the time-like 
and space-like
Killing vectors; so that the metric components do not
depend on coordinates
$t$ and $\varphi$, and the coordinates $\{r,\theta\}$ parametrize two-dimensional surfaces
orthogonal to the orbits of the symmetry group. Then we have:
\begin{eqnarray}
&\bm{g} = \gamma_{tt} \bm{\omega}^t{\otimes }\bm{\omega}^t
+\gamma_{t\varphi}(\bm{\omega}^t{\otimes }\bm{\omega}^\varphi+\bm{\omega}^\varphi{\otimes }\bm{\omega}^t)+
\gamma_{\varphi\varphi} \bm{\omega}^\varphi{\otimes }\bm{\omega}^\varphi
\nonumber\\
&\quad +  \gamma_{rr} \bm{\omega}^r{\otimes }\bm{\omega}^r+
\gamma_{r\theta}(\bm{\omega}^r{\otimes }\bm{\omega}^\theta+\bm{\omega}^\theta{\otimes }\bm{\omega}^r) 
+\gamma_{\theta\theta} \bm{\omega}^\theta{\otimes }\bm{\omega}^\theta ,
\label{eqmetrica}
\end{eqnarray}
where
$\bm{\omega}^t=dt$, $\bm{\omega}^r=dr$, $\bm{\omega}^\theta=r d\theta$, $\bm{\omega}^\varphi=r\sin\theta d\varphi$ 
is the Euclidean orthonormal co-basis associated with these coordinates.

Furthermore, coordinates $\{t, x=r\sin\theta\cos\varphi, y=r\sin\theta\sin\varphi, 
z=\cos\theta\}$ associated with the spherical-like coordinates are harmonic
and the metric in these
coordinates tends to the Min\-kows\-ki metric in standard Cartesian 
coordinates for large values of the coordinate $r$.

We assume that the source of the gravitational field is a perfect fluid,
\begin{equation}
\bm{T} = \left(\mu + p\right)\bm{u}\otimes\bm{u}+p \bm{g}
\label{eqenermom}
\end{equation}
whose density  and pressure $p$ are functions of the $r$ and
$\theta$ coordinates. Moreover, we assume the fluid has no convective motion, so its velocity
$\bm{u}$ lies on the plane spanned by the two Killing vectors,
\begin{equation}
\bm{u} = \psi\left(\bm\xi + \omega \bm\zeta\right) ,
\label{velocidad}
\end{equation}
where
\begin{equation}
\psi \equiv \left[-\left(\gamma_{tt}+2\omega \gamma_{t\varphi} r\sin\theta+\omega^2
 \gamma_{\varphi\varphi} r^2\sin^2\theta\right)\right]^{-\frac12}
\label{eqnorma}
\end{equation}
is a normalization factor, i.e., $u^\alpha u_\alpha=-1$.

Let us consider the Euler equations for the fluid (or the energy--momentum tensor conservation law, which is equivalent) for rigid motion we have:

\begin{equation}
\partial_a p = (\mu + p)\left( \partial_a\ln\psi\right)
\qquad (a,b,\dots = r ,\theta) .\label{eq5}
\end{equation}

For a barotropic fluid, $\mu(p)$ the integrability conditions for  (\ref{eq5}) are already satisfied.

Therefore the solution of equations (\ref{eq5}) is implicitly defined by the equation
\begin{equation}
 \int^p \frac{dp'}{\mu(p')+p'}=\ln\psi \label{Euler1}
\end{equation}
Since $p$ must be a function of  $\psi$ it will determine the surfaces $p=\mbox{constant}$ this surface can implicitly be defined as:
\begin{equation}
\psi=\psi_\Sigma ,
\label{eqsuperficie}
\end{equation}
where $\psi_\Sigma$ is an arbitrary constant.

Equation (\ref{Euler1}) and (\ref{eqsuperficie}) play an important role in our scheme. We use them to derive approximate expressions for the
pressure and the matching surface in a coherent way with the expansion for the metric we propose below.

Given an equation of state (EoS), we can integrate the left-hand side of (\ref{Euler1}) and even obtain explicit expressions for the pressure and density. For instance, a  linear equation of state,
$\mu+(1-n)p=\mbox{constant}$, with the condition of constant pressure on $\Sigma_1$ the surface which give the matching between the two fluid and if for the core we take $n=n_1$ and the constant equal to $\mu_1$ we obtain for the pressure and the density
\begin{equation}
 p=\frac{\mu_1}{n_1}\left(k\left(\frac{\psi}{\psi_{\Sigma_1}}\right)^{n_1}-1\right)\quad\mbox{and}\quad \mu=\frac{\mu_1}{n_1}\left(k(n_1-1)\left(\frac{\psi}{\psi_{\Sigma_1}}\right)^{n_1}+1\right)\label{eqmuandp1}
\end{equation}
When $\psi=\psi_{\Sigma_1}$ we have $p=\mu_1(k-1)/n_1$.

For the exterior layer we take  $n=n_2$, the constant equal to $\mu_2$,  and as the pressure in the exterior matching  surface must be zero and the constant value for the $\psi$ equal to $\psi_{\Sigma_2}$  then  we obtain
\begin{equation}
 p=\frac{\mu_2}{n_2}\left(\left(\frac{\psi}{\psi_{\Sigma_2}}\right)^{n_2}-1\right)\quad\mbox{and}\quad \mu=\frac{\mu_2}{n_2}\left((n_2-1)\left(\frac{\psi}{\psi_{\Sigma_2}}\right)^{n_2}+1\right)\label{eqmuandp2}
\end{equation}
 in the expressions  (\ref{eqmuandp1}), (\ref{eqmuandp2})  as the pressure must be the same computed from the core or from the crust then on the interior matching surface 
$$\frac{\mu_1}{n_1}\left(k-1\right)=\frac{\mu_2}{n_2}\left(\left(\frac{\psi_{\Sigma_1}}{\psi_{\Sigma_2}}\right)^{n_2}-1\right)$$
this equality determine the constant  $k$ in terms of the constants of  the EoS and the values of $\psi_\Sigma$ on both surfaces
\begin{equation}
k=1+\frac{\mu_2 n_1}{\mu_1 n_2}\left(\left(\frac{\psi_{\Sigma_1}}{\psi_{\Sigma_2}}\right)^{n_2}-1\right)\label{eqk1}
\end{equation}

\section{Approximation scheme}
As in our previous work \cite{CMMR,CuGMR,MMR,CuMMR} on the rigid rotation problem,  we introduce a post-Minkowskian dimensionless parameter, $\lambda=m/r_0$ where $m$ is the Newtonian mass, $r_0$ is the radius of the non rotating source. If $r_1$ is the radius of the interior core a good estimate of the total Newtonian mass is
\begin{equation}
m= \frac {4\pi}{3}\left(r_1^{3}\mu_1 + (
r_0^{3} - r_1^{3})\mu_2) \right)\label{Nm}
\end{equation}
That means that the mass and the density taking into account the definition of $\lambda$ begin to order $\lambda$ furthermore if we define the mean Newtonian density  as
$$ 3 m/(4\pi r_0^3)=3\lambda/(4\pi r_0^2)$$
we can redefine the density of the layers in terms of this mean density 
$$\mu_1= \varepsilon_1\frac{3\lambda}{4\pi r_0^2},\quad \mu_2= \varepsilon_2\frac{3\lambda}{4\pi r_0^2},$$
then  the equation (\ref{Nm}) leads to
\begin{equation}
1=\tau ^{3}\varepsilon_1 + (1 - \tau ^{3})\varepsilon_2\label{deftau}
\end{equation}
where $$\tau=\frac{r_1}{r_0},\quad 0<\tau<1,$$ 
is the ratio of the radius of the interior and exterior layers for the non-rotating source. The above expression (\ref{deftau}) means a relationship among the parameters densities and the radius of the core and the source radius.
Then in terms of $\varepsilon_1\mbox{ and }\varepsilon_2$, $\tau$ may be written
$$\tau=\left(\frac{1-\varepsilon_2}{\varepsilon_1-\varepsilon_2}\right)^{1/3}$$
And as in our previous work we introduce another dimensionless parameter, $\Omega=\lambda^{-1/2}\omega r_0$, that is a good estimate of the ratio between the gravitational and the rotational energies on the surface of the source.
 
Moreover, we have assumed the following expansion of the metric components, (see \cite{CMMR}) for the interior and exterior metrics  (we will not use labels to distinguish between exterior or interior metrics  whenever it can be clearly understood to which of them we are referring). 
\begin{eqnarray}
&&\gamma_{tt} \approx -1 + \lambda f_{tt} ,\quad
\gamma_{t\varphi} \approx\lambda^{3/2}\Omega f_{t\varphi} ,\quad
\gamma_{\varphi\varphi} \approx 1+\lambda f_{\varphi\varphi} ,
\nonumber\\
&&\gamma_{rr} \approx 1 + \lambda f_{rr} ,\quad 
\gamma_{r\theta} \approx\lambda f_{r\theta} ,\quad
\gamma_{\theta\theta} \approx 1+\lambda f_{\theta\theta}.\label{lambdaDepen}
\end{eqnarray}

With these metric behavior we obtain the following approximate expression for $\psi$
\begin{equation}
 \psi\approx 1+\lambda\frac12 \left(f_{tt}+\Omega^2\frac{\rho^2}{r_0^2}\right)\label{eqpsi}
\end{equation}
and we can also write $\psi_\Sigma=1+\lambda S_\Sigma$, if we substitute this expansions in the equations for the pressure and density (\ref{eqmuandp1}, \ref{eqmuandp2}) and in (\ref{eqk1}) we obtain that 
$$k=1+\lambda\frac{n_1\varepsilon_2}{\varepsilon_1}(S_1-S_2)+O(\lambda^2)$$ Vera, R.
with this dependences on $\lambda$ from equations (\ref{eqmuandp1}), (\ref{eqmuandp2})  we obtain that the density begin to order $\lambda$ and the pressure to order $\lambda^2$, i.e. to first order in  $\lambda$ the results for a linear EoS are the same that for a constant density EoS.

\section{First-order metric in harmonic coordinates}
As in our previous papers \cite{CMMR,CuGMR,MMR,CuMMR}, here we use the post-Minkowskian approximation scheme.
\begin{equation}
g_{\alpha \beta}=\eta_{\alpha \beta}+\lambda h_{\alpha \beta}
\end{equation}
In those references, the resulting equations and notation are explained.
\subsection{Linear exterior solution}
The inhomogeneous part of the linear exterior equations is zero, i.e.:
$$t_{\alpha\beta}=0$$ 
and the equations to solve are:
\begin{eqnarray}
&&\triangle h_{\alpha\beta} = 0 ,\nonumber\\
[.6ex]
&&\partial^k (h_{k\mu} -\frac12
h \eta_{k\mu} ) = 0 .
\label{eqhomog}
\end{eqnarray}
We are going to assume equatorial symmetry and the same dependence of the metric on the expansion parameters  as in our previous work \cite{CMMR,CuGMR} ($M_n\propto \Omega^n$ , $J_n\propto \Omega^n$). So, the exterior metric up to order $\lambda$ and $\Omega^3$ 
can be written in terms of the spherical harmonic tensors as:
\begin{equation}
\bm{h}\approx 2\lambda\sum_{l=0,2}\Omega^l\frac{M_l}{\eta^{l+1}}\left(\bm{T}_l+\bm{D}_l\right)
+2\lambda^{3/2}\sum_{l=1,3}\Omega^l\frac{J_l}{\eta^{l+1}} \bm{Z}_l
\label{eqsolinexthar}
\end{equation}
where $\eta\equiv r/r_0$
\begin{eqnarray}
&&\bm{T}_n  \equiv P_n(\cos\theta) \bm{\omega}^t\otimes \bm{\omega}^t \quad (n\geq 0) ,
\nonumber\\
&&\bm{D}_n \equiv P_n(\cos\theta) \delta_{ij}dx^i{\otimes }dx^j \quad (n\geq 0) ,
\nonumber\\
&&\bm{Z}_n  \equiv
P_n^1(\cos\theta) (\bm{\omega}^t\otimes\bm{\omega}^\varphi+\bm{\omega}^\varphi\otimes\bm{\omega}^t)
\quad (n\geq 1) ,
\label{base1}
\end{eqnarray}
are spherical harmonic tensors, and $P_n$ are the Legendre polynomials and $P^m_n$ the associated Legendre functions.

We have not written the gauge terms  because to this order in $\lambda$ are not needed for the matching.  The gauge terms for the interior or exterior solution define different harmonic coordinates sets and are only needed when we match the interior with the exterior metric, but for this particular solution to first order they are zero, and for this reason we do not write them here.

Let us remark that in this approximation order $M_0$ and $J_1$ can be polynomials of first order in $\Omega^2$ and $M_2$, $J_3$ are pure numbers.

To obtain the exterior metric, we must add the Minkowski part to the exterior solution for $\bm{h}$ (\ref{eqsolinexthar}), i.e.:
\begin{equation}
\bm{g}_{\rm ext}\approx -\bm{T}_0+\bm{D}_0+\bm{h}.\label{extg1}
\end{equation}
\subsection{Linear interior solution}
We will now find the interior solution for a fluid with linear EoS, in this problem we have two fluids with different densities.
To this end, we need an approximate expression of the energy--momentum tensor of the fluid in both layers. Therefore, the 
energy--momentum tensor (\ref{eqenermom}) contributes to the right-hand side of the Einstein equations by means of:
\begin{equation}
8\pi\,\bm{t} \approx\varepsilon\left(
3\frac{\lambda}{ r_0^2}\left(\bm{T}_0+\bm{D}_0\right)
+6\frac{\lambda^{3/2}\Omega}{r_0^2}\eta\,\bm{Z}_1\,,\right)
\label{impulsenerg0}
\end{equation}
if the terms of order equal to or higher than $\lambda^2$ are disregarded. 

Let us consider the following system of linear differential equations that corresponds to the linear post--Minkowskian approximation:
\begin{eqnarray}
&&\triangle h_{\alpha\beta} = -16\pi t_{\alpha\beta}\,,
\nonumber\\[.6ex]
&&\partial^k (h_{k\mu} -\frac12 h\,\eta_{k\mu}\,) = 0\,,
\label{eqint}
\end{eqnarray}
where $\bm{t}$ is given by (\ref{impulsenerg0}) with $\epsilon\rightarrow\epsilon_2$.
A particular solution for the inhomogeneous part is:
\begin{equation}
 h_{\rm inh2} = \varepsilon_2\left(- \lambda ^{3/2}\frac{6\Omega}{5}\eta^3 \bm{Z}_1
-\eta^2(\bm{T}_0+\bm{D}_0)\right)\label{inh2}
\end{equation}
We can add a solution of the homogeneous equation with equatorial symmetry which in this case can be written up to order $\lambda$ and $\Omega^3$ as
\begin{eqnarray}
&& \hspace*{-3.5em}h_{\rm hom2} =\lambda ^{3/2}\Omega \left(\left(\mathfrak{j}_1\,   \eta +\frac{2\,\mathfrak{J}_1}{\eta^2}\right)\bm{Z}_1+
 \Omega^2\left(\mathfrak{j}_3\eta^3 
 +   \frac {2 \mathfrak{J}_3}{\eta^4}\right) \bm{Z}_3\right)+\nonumber \\ &&  
 \left( \mathfrak{m}_0 + \frac {2 \mathfrak{M}_0}{\eta}\right)(\bm{T}_0+\bm{D_0})+\Omega ^2\left(\mathfrak{m}_2 \eta^2+\frac{2 \mathfrak{M}_2}{\eta^3}\right)(\bm{T}_2+\bm{D_2})
\end{eqnarray}
The main difference between this and our previous papers is that the homogeneous part contain terms that depend on $1/r$ because they are not singular in this layer, and also we have skipped all the gauge terms because they are not needed for the matching to first order.

As before, in  this approximation order, $\mathfrak{m}_0$, $\mathfrak{M}_0$, $\mathfrak{j}_1$ and $\mathfrak{J}_1$ are linear functions of $\Omega^2$ and  $\mathfrak{m}_2$, $\mathfrak{M}_2$ $\mathfrak{j}_3$ $\mathfrak{J}_3$  are pure numbers

Finally, adding this homogeneous part to the inhomogeneous part, and to the Minkowski part, we obtain an approximate expression for the  metric of the layer
up to the order $\lambda^{3/2}$ and $\Omega^3$:
\begin{equation}
\bm{g}_{\rm int2}\approx -\bm{T}_0+\bm{D}_0+\bm{h}_{\rm hom2}+\bm{h}_{\rm inh2}. \label{intg2}
\end{equation}

To  find the core metric we proceed as before now the energy momentum tensor is also given in (\ref{impulsenerg0}) but with $\varepsilon\rightarrow \varepsilon_1$ then a particular solution is $h_{\rm inh2}$ given in eq.(\ref{inh2}) with $\varepsilon_2\rightarrow \varepsilon_1$ which  we denote as $h_{\rm inh1}$. The homogeneous part now is shorter because the terms with $1/\eta$ are singular at the origin and then are not present, we can write for the homogeneous solution of the interior metric  of the core where we have also skipped all the gauge terms which are not needed for the matching to first order
\begin{equation}
h_{\rm hom1} =\lambda ^{3/2}\Omega \left(j_1\,   \eta \bm{Z}_1+
 \Omega^2 j_3\eta^3 
  \bm{Z}_3\right)+ 
 m_0 (\bm{T}_0+\bm{D_0})+\Omega ^2 m_2 \eta^2(\bm{T}_2+\bm{D_2})
\end{equation}
finally the metric of the core is
\begin{equation}
\bm{g}_{\rm int1}\approx -\bm{T}_0+\bm{D}_0+\bm{h}_{\rm hom1}+\bm{h}_{\rm inh1}. \label{intg1}
\end{equation}

\subsection{Matching  surface}
If we assume that the metric components are continuous on the matching surfaces, then we can use their exterior expressions for the exterior matching surface given by (\ref{extg1}) and (\ref{eqpsi}) to make (\ref{eqsuperficie}) into  a true equation. So
we can search for a parametric form of the matching surface up to zero order in $\lambda$ and up to order $\Omega^2$ by making the following assumption:
\begin{equation}
r\approx r_0\left(1+\sigma_2 \Omega^2 P_2(\cos\theta)\right)\,.
\label{desenvradisup0}
\end{equation}
where $\sigma_2$  must be determined from the equation $(\ref{eqsuperficie})$. 
The function $f_{tt}$ can be read from  (\ref{extg1}), i.e.:
$$f_{tt}=2 \sum_{n=0,2} \Omega^n\frac{M_n}{\eta^{n+1}}P_n(\cos \theta)$$
A simple calculation leads to:
\begin{equation}
\sigma_2=\frac{3M_2-1}{3M_0}
\label{sigma2}
\end{equation}

We can also obtain a similar expression for  $\psi_{\Sigma_2}$ (\ref{eqsuperficie}) in terms of the exterior constants:
\begin{equation}
\psi_{\Sigma_2}\approx 1+\lambda\left(M_0 +\frac{1}{3}\Omega^2\right)\,.
\label{psis2}
\end{equation}
We can also compute this expression in terms of the $f_{tt}$ of the layer  but this gives an involved expression in terms of the layer constants, but once the Lichnerovicz matching is performed (i.e. the continuity of the metric) both expressions give  the same result.

For the interior surface we use  the same procedure
\begin{equation}
r\approx r_0\tau\left(1+\sigma_1 \Omega^2 P_2(\cos\theta)\right)\,.
\label{desenvradisup1}
\end{equation}
and now we compute the expression in terms of the $f_{tt}$ of the  core 
\begin{equation}
f_{tt}=m_0 + \frac{m_2\Omega^2r^2P_2(\cos\theta)}{r_0^2}- 
\frac {\varepsilon _1r^2}{r_0^2}
\end{equation}
and we obtain for $\sigma_1$
\begin{equation}
\sigma_1=\frac{3m_2 - 2}{6\varepsilon_1} 
\label{sigma1} 
\end{equation}
as in the previous case we can also compute this expression in terms of the $f_{tt}$ of the layer and the same result is obtained after the matching of the core and the layer. For  $\psi_{\Sigma_1}$ we obtain
\begin{equation}
\Psi_{\Sigma_1}\approx 1+\lambda\left(\frac{m_0}{2}+\left(\frac{\Omega ^2}{3} - \frac{\varepsilon_1}{2}\right)\tau ^{2} \right)
\end{equation}
\subsection{Global solution to first order in $\lambda$}
Let us recall the matching conditions we are using in this paper:  the metric components and their first derivatives have to be continuous.
Imposing these conditions on the metrics $\bm{g}_{\rm ext}$ and $\bm{g}_{\rm int2}$ through the  hyper-surface   $\Sigma_2$ of zero pressure, and on the metrics $\bm{g}_{\rm int1}$ and $\bm{g}_{\rm int2}$ on the  interior surface,  $\Sigma_1$, of constant pressure and  bearing in mind the order of approximation we are concerned with we obtain for the constants the following expressions for the core constants
\begin{eqnarray}
&&m_0=3\varepsilon_2 + 3\tau^2(\varepsilon_1 -\varepsilon_2), \quad m_2=\frac{2}{3}\left(1-25\frac{\varepsilon_1}{d}\right),\nonumber\\[.6ex]
&&j_1=2\varepsilon_2 + 2\tau ^2(\varepsilon_1 - \varepsilon_2) + 
 \frac{4\Omega^2}{9}\left( - 1 +\frac{5(3\tau ^2(\varepsilon_1 - \varepsilon_2) + 3\varepsilon_2 + 2\varepsilon_1)}{d}\right),\nonumber\\[.6ex] 
 && j_3=\frac {4}{21}\left(1  -25\frac{\varepsilon_1}{d}\right)\label{interior_cons}
 \end{eqnarray}
 where
 $$d\equiv (5-3\varepsilon_2)(3\varepsilon_2 +2\varepsilon_1)-9\varepsilon_2( 1-\varepsilon_2)\tau ^2$$
 and for the layer constants we obtain 
 \begin{eqnarray}
 &&\mathfrak{M}_0=1-\varepsilon_2, \quad \mathfrak{m}_0=3\varepsilon_2,\quad\mathfrak{m}_2=\frac{2}{3}\left(1 -\frac{5}{d}(3\varepsilon_2+2\varepsilon_1)\right),\nonumber\\[.6ex]
&&\mathfrak{M}_2=-\frac{5}{d}(1-\varepsilon_2)\tau ^2,\quad \mathfrak{j}_1= 2\varepsilon_2+\frac {4}{9}\Omega^2\left(-1 +\frac{5}{d}(3\varepsilon_2+2\varepsilon_1)\right) ,\nonumber\\[.6ex]
 &&\mathfrak{J}_1= (1 - \varepsilon_2)\tau^2\left(\frac{2}{5}+\frac {10\Omega^2}{3d}\right), \quad\mathfrak{j}_3= \frac{4}{21}\left(1 -\frac{5}{d}(3\varepsilon_2+2\varepsilon_1)\right) ,\nonumber\\[.6ex]
&&\mathfrak{J}_3=-\frac {10}{7}\frac{\tau^4}{d}(1 - \varepsilon_2), \label{layer_cons}
 \end{eqnarray}
 and finally for the multipole moments 
\begin{eqnarray}
&&M_0=1,\quad M_2= \frac {1}{3}  + \frac {5}{9\varepsilon_2}\left(1 -\frac{5}{d}(3\varepsilon_2 + 2\varepsilon_1)\right),\nonumber\\[.6ex]
&&J_1=\frac{2}{5}\left(\varepsilon_2+(1 - \varepsilon_2)\tau ^2\right)  +  \frac{2\Omega^2}{9}\left(- 1 + \frac{5(3\tau^2(1 - \varepsilon_ 2) + 3\varepsilon_2 +2\varepsilon_1)}{d}\right),\nonumber\\[.6ex]
 && J_3= \frac {2}{21} \left(1-\frac{5}{d} \left( 3\varepsilon_2 + 2\varepsilon_1+3\tau^4(1 - \varepsilon_2) \right)\right)\label{exterior_cons}
\end{eqnarray}

\section{Conclusions}
In this paper we have obtained a global approximate slow rotating solution in harmonic coordinates for a double layer star matched on a zero pressure surface to an asymptotically flat exterior, the solution depend only on the constants of the two linear  equations of state $\epsilon_1$ and $\epsilon_2$, the radius of the matching surface $r_0$ for the non rotation limit and the constant angular velocity $\Omega$.
 
First of all, let us notice that  the densities
are constant quantities of order 
$\lambda$ so, to this order of approximation in $\lambda$, the density constant EoS and the linear one give the same energy momentum tensor. But once obtained  the metric to order $\lambda$ we can obtain the densities and pressures to order $\lambda^2$ and so the energy-momentum tensor up to order $\lambda^2$.

The pressure in the crust is
\begin{eqnarray}
p&=&\frac{6\lambda^2}{r_0^2}\epsilon_2\left\{\frac32\epsilon_2-1-\frac{\Omega^2}{3}-\left[\frac{\epsilon_2}{2}-\frac{\Omega^2}{3}\left(1-5\frac{3\epsilon_2+2\epsilon_1}{d}P_2(\cos(\theta))\right)\right]\eta^2+
\right.\nonumber\\ && \left. (1-\epsilon_2)\frac1\eta-5\Omega^2(1-\epsilon_2)\frac{\tau^2}{d}P_2(\cos(\theta))\frac{1}{\eta^3}\right\}\label{pressure2}
\end{eqnarray}
and the pressure into the core is
\begin{eqnarray}
p&=&\frac{6\lambda^2}{r_0^2}\left\{
\epsilon_1\left(
\tau^2\left( \frac{\epsilon_1}{2}-\frac{\Omega^2}{3}\right)-\left[\frac{\epsilon_1}{2}-\frac{\Omega^2}{3}\left(1-25\frac{\epsilon_1}{d}P_2(\cos(\theta))\right)\right]\eta^2\right)+
\right. \nonumber \\ && \left. \epsilon_2\left(\tau^2\left( \epsilon_1+\frac{\Omega^2}{3}\right)-1- \frac{\Omega^2}{3}+\frac32 \epsilon_2(1-\tau^2)\right)\right\}\label{pressure1}
\end{eqnarray}
Another interesting fact is that with these results we  can also obtain the results for only one layer i.e. for an homogeneous distribution of matter in two different ways first by eliminating the exterior layer that is $\tau=1, \epsilon_1=1$ then we obtain the results of \cite{CMMR} for the constants of (\ref{interior_cons}) and (\ref{exterior_cons}) but also can be done by eliminating the interior layer, i.e $\tau=0, \epsilon_2=1$ then we obtain for the constants of (\ref{layer_cons}) and (\ref{exterior_cons}) the same result and what is interesting to note is that the singular part in (\ref{layer_cons}) vanishes as it would be,  this is so because these constants are determined by doing the matching with a regular interior.

We can also obtain the metric and pressure for $\epsilon_1=0$  i.e. for an EoS $\mu+(1-n)p=0$. In this case pressure and density begin to order $\lambda^2$ and from  (\ref{pressure1}) we can see that to this order the pressure into the core is constant.

There is no problem to go further to the next order, in fact we have obtained the solution for the two layers model up to order $\lambda^{5/2}$ and $\Omega^3$  but the expressions are so cumbersome that we do not  write them in this paper. 

\section*{Acknowledgements}
AM gratefully acknowledge the \textit{Universidad de Salamanca }for the warm hospitality which facilitated this collaboration. Financial support to the authors for this work was provided by the Spanish people via the government award FIS2015-65140-P (MIN\-E\-CO/FE\-DER).

\end{document}